\documentclass[aps,prd,onecolumn,groupedaddress,showpacs,nofootinbib,amssymb]{revtex4}
\usepackage[T1]{fontenc}
\usepackage[latin1]{inputenc}
\usepackage{graphicx}
\usepackage[english]{babel}
\usepackage{amsmath}
\usepackage{amssymb}
\usepackage{amsfonts}

\begin{document}
\def\pp{{\, \mid \hskip -1.5mm =}}
\def\cL{{\cal L}}
\def\be{\begin{equation}}
\def\ee{\end{equation}}
\def\bea{\begin{eqnarray}}
\def\eea{\end{eqnarray}}
\def\beq{\begin{eqnarray}}
\def\eeq{\end{eqnarray}}
\def\tr{{\rm tr}\, }
\def\nn{\nonumber \\}
\def\e{{\rm e}}

\title{\textbf{$f(R)$ gravity constrained by PPN parameters and stochastic background of gravitational waves }}

\author{S. Capozziello$^1$, M. De Laurentis$^2$$^1$, S. Nojiri$^3$, S. D. Odintsov$^4$}

\affiliation{\it $^1$Dipartimento di Scienze fisiche, Università
di Napoli {}`` Federico II'', INFN Sez. di Napoli, Compl. Univ. di
Monte S. Angelo, Edificio G, Via Cinthia, I-80126, Napoli, Italy\\
$^2$Politecnico di Torino and INFN Sez. di
Torino, Corso Duca degli Abruzzi 24, I-10129 Torino, Italy\\
$^3$ Department of Physics, Nagoya University, Nagoya 464-8602, Japan\\
$^4$Institucio Catalana de Recerca i Estudis Avancats (ICREA) and
Institut de Ciencies de l Espai (IEEC-CSIC), Campus UAB, Facultat
de Ciencies, Torre C5-Par-2a pl, E-08193 Bellaterra (Barcelona),
Spain.}

\date{\today}

\begin{abstract}
We analyze seven different viable $f(R)$-gravities towards the
Solar System tests and stochastic gravitational waves background.
The aim is to achieve experimental bounds for the theory at local
and cosmological scales in order to select models capable of
addressing the accelerating cosmological expansion without
cosmological constant but evading the weak field constraints.
Beside large scale structure and galactic dynamics, these bounds
can be considered complimentary in order to select self-consistent
theories of gravity working at the infrared limit. It is
demonstrated that seven viable $f(R)$-gravities under
consideration not only satisfy the local tests, but additionally,
pass the above PPN-and stochastic gravitational waves bounds for
large classes of parameters.
\end{abstract}


\maketitle


\vspace{5.mm}
\section{Introduction}
The currently observed accelerated expansion of the Universe
suggests that cosmic flow dynamics is dominated by some unknown
form of dark energy characterized by a large negative pressure.
This picture comes out when such a new ingredient, beside baryonic
and dark matter,  is considered as a source in the r.h.s. of the
field equations. Essentially, it should be some form of
un-clustered, non-zero vacuum energy which, together with
(clustered) dark matter, should drive the global cosmic  dynamics.

Among the proposals to explain the experimental situation,  the
``concordance model'', addressed as $\Lambda$CDM,  gives a
reliable snapshot of the today observed Universe according to the
CMBR, LSS and SNeIa data,  but presents dramatic shortcomings as
the ``{\it coincidence and  cosmological constant problems}''
which point out its inadequacy  to fully trace back the
cosmological dynamics \cite{Peebles}.

On the other hand, alternative theories of gravity,  extending in
some way General Relativity (GR), allows to pursue a different
approach  giving rise to suitable cosmological models where a
late-time accelerated expansion can be achieved in several  ways.
This viewpoint does not require to find out candidates for dark
energy and dark matter at fundamental level (they have not been
detected up to now), it takes into account only the ``observed''
ingredients (i.e. gravity, radiation and baryonic matter), but the
l.h.s. of the Einstein equations has to be modified.  Despite of
this modification, it could be in agreement with the spirit of GR
since the only request is that the Hilbert-Einstein action should
be generalized asking for a gravitational interaction  acting, in
principle,  in different ways at different scales \cite{CCT}.

The idea that Einstein gravity should be extended or corrected at
large scales (infrared limit) or at high energies (ultraviolet
limit) is suggested by several theoretical and observational
issues. Quantum field theory in curved spacetimes, as well as the
low-energy limit of String/M theory, both imply semi-classical
effective actions containing  higher-order curvature invariants or
scalar-tensor terms. In addition, GR has been definitely tested
only at Solar System scales while it may show several shortcomings
if checked at higher energies or larger scales. Besides,  the
Solar System experiments are, up to now, not so conclusive to
state that the only viable theory of gravity  is GR: for example,
the limits on PPN parameters should be greatly improved to fully
remove degeneracies \cite{gaia}.

Of course, modifying the gravitational action asks for several
fundamental challenges. These models can exhibit instabilities
\cite{instabilities-f(R)} or ghost\,-\,like behavior
\cite{ghost-f(R)}, while, on the other hand, they have to be
matched with observations and experiments in the appropriate low
energy limit.

Despite of all these issues, in the last years, some interesting
results have been achieved in the framework of the so called
$f(R)$-gravity at cosmological, Galactic and Solar System scales.
Here $f(R)$ is a general (analytic)  function of the Ricci scalar
$R$ (see Refs.~\cite{odirev,GRGrev,faraoni} for review).

For example, there exist cosmological solutions that give the
accelerated expansion of the universe at late times
\cite{f(R)-noi,f(R)-cosmo,mimic,prado}.  In addition, it has been
discovered that some stability conditions can lead to avoid ghost
and tachyon solutions. Furthermore there exist viable $f(R)$
models which satisfy both background cosmological constraints and
stability conditions
\cite{Li,Hu,Star,Appleby,Tsuji,Nojiri:2007cq,sebast,Nojiri} and
results have been achieved in order  to place constraints on
$f(R)$ cosmological models by CMBR anisotropies and galaxy power
spectrum \cite{bean-f(R),Faul,huspe}. Moreover, some of such
viable models lead to the unification of early-time inflation with
late-time acceleration \cite{Nojiri:2007cq,sebast,Nojiri}.

On the other hand, by considering $f(R)$-gravity in the low energy
limit, it is possible to obtain corrected gravitational potentials
capable of explaining  the flat rotation curves of spiral galaxies
or the  dynamics of galaxy clusters without considering huge
amounts of dark matter
\cite{noi-mnras,salucci,sobouti,mendoza,Boe,lobo}.

Furthermore, several authors have dealt with the weak field limit
of fourth order gravity, in particular considering the PPN limit
\cite{bertolami,ppn-tot1,ppn-tot,Navarro,Erick,lgcpapers,mpla,CST}
and the spherically symmetric solutions
\cite{spher-f(R),multamaki06,kainulainen,noether}.

This great deal of work needs an essential issue to be pursued: we
need to compare experiments and probes at local scales (e.g. Solar
System) with experiments and probes at large scales (Galaxy,
extragalactic scales, cosmology) in order to achieve
self-consistent $f(R)$ models. Some work has been done in this
direction (see e.g. \cite{Hu}) but the large part of efforts has
been devoted to address single data sets (observations at a given
redshift) by a single model which, several time, is not working at
other scales than the one considered. In particular, a given
$f(R)$ model, evading Solar System tests, should be not simply
extrapolated at extragalactic and cosmological scales only
requiring accelerated cosmological solutions but it should be
confronted with data and probes coming from cosmological
observations. Reliable models are then those matching data at very
different scales (and redshifts).

In order to constrain further viable $f(R)$-models, one could take
into account also  the stochastic background of gravitational
waves (GW) which, together with cosmic microwave background
radiation (CMBR), would  carry a huge amount of information on the
early stages of the Universe evolution.  In fact, if detected,
such a background could constitute a further probe for these
theories at very high red-shift \cite{tuning}. On the other hand,
a key role for the production and the detection of the relic
gravitational radiation background is played by the adopted theory
of gravity \cite{Maggiore,BBFN}. This means that the effective
theory of gravity should be probed at zero, intermediate and high
redshifts to be consistent at all scales and not simply
extrapolated up to the last scattering surface, as in the case of
GR.

The aim of this paper is to discuss the  PPN Solar-System
constraints and the GW stochastic background considering some
recently proposed $f(R)$ gravity models
\cite{Star,Li,Hu,Nojiri:2007cq,sebast,Nojiri} which satisfy both
cosmological
and stability conditions mentioned above. Using the definition of
PPN-parameters $\gamma$ and $\beta$ in terms of $f(R)$-models
\cite{mpla} and the definition of scalar GWs \cite{CCD}, we
compare and discuss if it is possible to search for parameter
ranges of  $f(R)$-models  working at Solar System and GW
stochastic background scale. This phenomenological approach is
complementary to the one proposed, e.g. in \cite{Hu,huspe} where
also galactic and cosmological scales have been considered to
constraint the models.

The layout of the paper is the following.  In  Sec. II, we review
the field equations of $f(R)$ gravity in the metric approach and
their scalar-tensor representation, useful to compare the theory
with observations. In Sec.III, we review and discuss some viable
$f(R)$ models capable of satisfying both local gravity
prescriptions  as well as the observed cosmological behavior. In
particular, we discuss their stability conditions and the field
values which have to achieved to fulfill physical bounds. Sec. IV
is devoted to derive the values of model parameters in agreement
with the PPN experimental constraints while, in Sec. V, we deal
with the constraints coming from the stochastic background of GWs.
These latter ones have to be confronted with those coming from PPN
parameterization. Discussion and conclusions are drawn in Sec. VI.
As a general remark, we find out that bounds coming from the
interferometric ground-based (VIRGO, LIGO) and space (LISA)
experiments could constitute a further probe for $f(R)$ gravity if
matched with bounds at other scales.

\section{ $f(R)$ gravity}

Let us start from the following action
\begin{equation}
\mathcal{S}=\mathcal{S}_g + \mathcal{S}_m =\frac{1}{k^{2}}\int
d^{4}x\sqrt{-g}\left[R+f(R)+\mathcal{L}_{m}\right]\,,\label{eq:action}\end{equation}
where we have considered the gravitational and matter
contributions and $k^2\equiv 16\pi G$. The non-linear $f(R)$ term
has been put in evidence with respect to the standard
Hilbert-Einstein term $R$ and $\mathcal{L}_{m}$ is the
perfect-fluid matter Lagrangian. The field equations are
\begin{equation}
\frac{1}{2}g_{\mu\nu}F(R)-R_{\mu\nu}F'(R)-g_{\mu\nu}\square F'(R)+
\nabla_{\mu}\nabla_{\nu}F'(R)=-\frac{k^{2}}{2}T_{\mu\nu}^{(m)}.\label{eq:motion}\end{equation}
Here $F(R)=R+f(R)$ and $T_{\mu\nu}^{(m)}$is the matter
energy\,-\,momentum tensor. By introducing the auxiliary field
$A$, one can rewrite the gravitational part in the Action
(\ref{eq:action}) as

\begin{equation}
\mathcal{S}_g=\frac{1}{k^{2}}\int d^{4}x\sqrt{-g}\left\{
\left(1+f'(A)\right)\left(R-A\right)+A+f(A)\right\}
.\label{eq:actionA}\end{equation} As it is clear from
Eq.(\ref{eq:actionA}), if $F'(R)=1+f'(R)<0$, the coupling
$k_{eff}^{2}=k^{2}/F'(A)$ becomes negative and the theory enters
the anti-gravity regime. Note that it is not the case for the
standard GR.

Action (\ref{eq:actionA}) can  be recast in a  scalar-tensor form.
By using the conformal scale transformation $g_{\mu\nu}\rightarrow
e^{\sigma}g_{\mu\nu}$ with $\sigma=-\ln\left(1+f'(A)\right)$, the
action can be written in the Einstein frame as follows
\cite{odirev}:

\begin{equation}
\mathcal{S}_{E}=\frac{1}{k^{2}}\int d^{4}x\sqrt{-g}\left(R-\frac{3}{2}g^{\rho\sigma}\partial_{\rho}\sigma
\partial_{\sigma}\sigma-V(\sigma)\right),\label{eq:actionE}\end{equation}

where

\begin{equation}
V(\sigma)=e^{\sigma}g\left(e^{-\sigma}\right)-e^{2\sigma}f\left(g\left(e^{-\sigma}\right)
\right)=\frac{A}{F'(A)}-\frac{F(A)}{F'(A)^{2}}.\label{eq:potential}\end{equation}

The form of  $g\left(e^{-\sigma}\right)$ is given by solving
$\sigma=-\ln\left(1+f'(A)\right)=\ln F'(A)$ as
$A=g\left(e^{-\sigma}\right)$. The transformation
$g_{\mu\nu}\rightarrow e^{\sigma}g_{\mu\nu}$ induces a coupling of
the scalar field $\sigma$ with  matter.

In general,  an effective mass for $\sigma$ is defined as
\cite{Nojiri}

\begin{equation}
m_{\sigma}^{2}\equiv\frac{1}{2}\frac{d^{2}V(\sigma)}{d\sigma^{2}}
=\frac{1}{2}\left[
\frac{A}{F'(A)}-\frac{4F(A)}{(F'(A))^{2}}+\frac{1}{F''(A)}\right]
\,, \label{eq:mass}\end{equation} which, in the weak field limit,
could induce corrections to the Newton law. This allows, as it is
well known, to deal with the extra degrees of freedom of
$f(R)$-gravity as an effective scalar field which reveals
particularly useful in considering "chameleon" models
\cite{chameleon}. This "parameterization" will be particularly
useful to deal with the scalar component of GWs.

\section{$f(R)$ viable models}

Let us  consider now a class of $f(R)$ models which do not contain
cosmological constant and  are explicitly designed to satisfy
cosmological and Solar-System constraints in given limits of the
parameter  space. In practice, we choose a class of functional
forms of $f(R)$ capable of matching, in principle, observational
data (see \cite{mimic} for the general approach). Firstly, the
cosmological model  should reproduce the  CMBR constraints in the
high-redshift regime (which agree with the presence of an
effective cosmological constant). Secondly, it should give rise to
an accelerated expansion, at low redshift, according to the
$\Lambda$CDM model. Thirdly, there should be sufficient degrees of
freedom in the parameterization to encompass low redshift
phenomena (e.g. the large scale structure) according to the
observations \cite{huspe}. Finally, small deviations from GR
should be consistent with Solar System tests. All these
requirements suggest that we can assume the  limits
\begin{equation}
\lim_{R\rightarrow\infty}f(R)={\rm constant},
\end{equation}
\begin{equation}
\lim_{R\rightarrow0}f(R)=0,
\end{equation}
which are satisfied by a general class of broken power law models,
proposed in \cite{Hu}, which are

\begin{equation}
f_{I}(R)=-m^{2}\frac{c_{1}\left(\frac{R}{m^{2}}\right)^{n}}{c_{2}\left(\frac{R}{m^{2}}\right)^{n}+1}
\label{eq:HS}\end{equation} or otherwise written as

\begin{equation}
F_{I}(R)=R-\lambda
R_{c}\frac{\left(\frac{R}{R_{c}}\right)^{2n}}{\left(\frac{R}{R_{c}}\right)^{2n}+1}
\label{eq:HS1}\end{equation}
where $m$ is a  mass scale and
$c_{1,2}$ are dimensionless parameters.

Besides, another viable class of models  was proposed in
\cite{Star}

\begin{equation}
F_{II}(R)=R+\lambda
R_{c}\left[\left(1+\frac{R^{2}}{R_{c}^{2}}\right)^{-p}-1\right]\,.\label{eq:STAROBINSKY}\end{equation}
Since $F(R=0)=0$, the cosmological constant has to disappear in a
flat spacetime. The parameters  $\{n$, $p$, $\lambda$, $R_{c}\}$
are constants which should be determined by experimental bounds.

Other interesting models with similar features  have been studied
in \cite{Nojiri,Nojiri:2007cq,Appleby,Tsuji,sebast}. In all these
models, a de-Sitter stability point, responsible for the late-time
acceleration, exists for $R=R_1~(>0)$, where $R_1$ is derived by
solving the equation $R_1 f_{,R}(R_1)=2f(R_1)$ \cite{ottewill}.
For example, in the model (\ref{eq:STAROBINSKY}), we have
$R_1/R_c=3.38$ for $\lambda=2$ and $p=1$. If $\lambda$ is of the
unit order, $R_1$ is of the same order of $R_c$. The stability
conditions, $f_{,R}>0$ and $f_{,RR}>0$, are fulfilled for $R>R_1$
\cite{Star,Tsuji}. Moreover the models satisfy the conditions for
the cosmological viability that gives rise to the sequence of
radiation, matter and accelerated epochs \cite{Tsuji}.

In the region $R \gg R_c$ both classes of models (\ref{eq:HS}) and
(\ref{eq:STAROBINSKY}) behave as
\begin{eqnarray}
\label{eq:fR} F_{III}(R) \simeq R-\lambda R_c \left[1- \left( R_c/R
\right)^{2s} \right]\,,
\end{eqnarray}
where $s$ is a positive constant. The model approaches
$\Lambda$CDM  in the limit $R/R_c \to \infty$.

\begin{figure}[ht]
\includegraphics[scale=0.6]{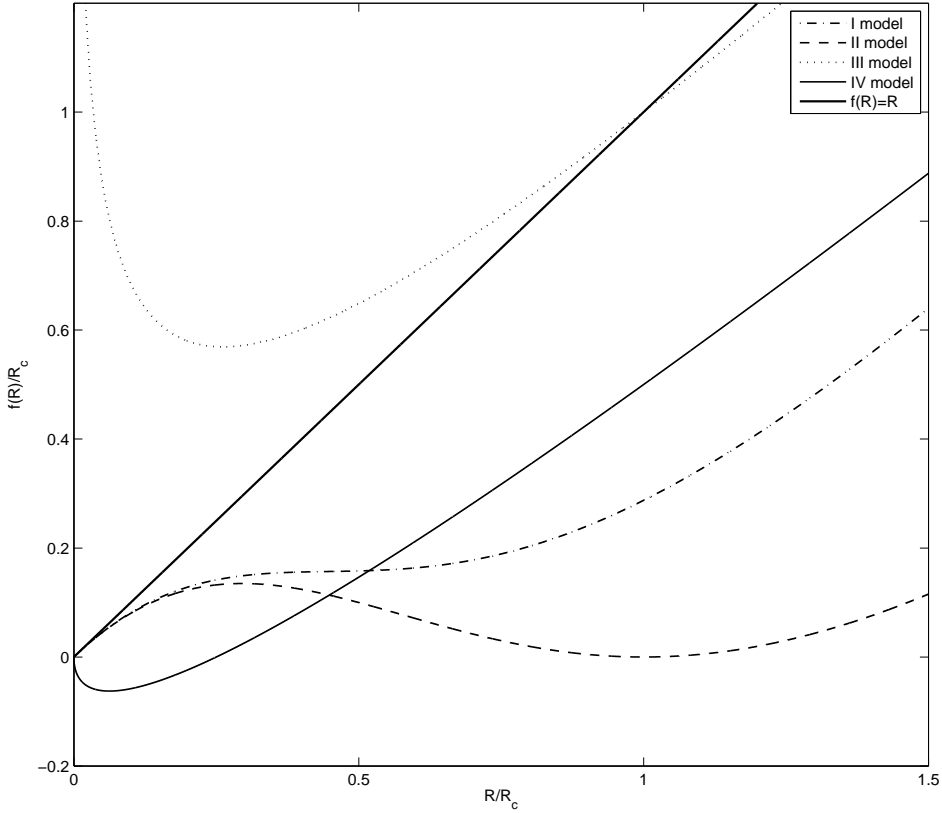}
\caption{Plots of  four different $F(R)$ models as function of
$\frac{R}{R_{c}}$.  Model I in Eq. (\ref{eq:HS}) with $n=1$ and
$\lambda=2$ (dashed line). Model II in Eq.(\ref{eq:STAROBINSKY})
with $p=2$, $\lambda=0.95$ (dashdot line). Model III in
Eq.(\ref{eq:fR}) with $s=0.5$ and $\lambda=1.5$ (dotted). Model IV
in   Eq.(\ref{eq:TS}) with $q=0.5$ and $\lambda=0.5$ (solid line).
We also plot  $F(R)=R$ (solid thick line) to see whether or not
the stability condition $F_{,R}>0$ is violated. } \label{fig:f(R)}
\end{figure}

Finally, let  also consider the class of models
\cite{Li,AT07,Faul}

\begin{equation}
F_{IV}(R)=R-\lambda
R_{c}\left(\frac{R}{R_{c}}\right)^{q}\,.\label{eq:TS}
\end{equation}
Also in this case $\lambda$, $q$ and $R_c$ are positive constants
(note that $n$, $p$, $s$ and $q$ have to converge toward the same
values to match the observations). We do not consider the models
whit negative $q$, because they suffer for instability problems
associated with negative $F_{,RR}$ \cite{DK,bean-f(R)}. In
Fig.(\ref{fig:f(R)}), we have plotted some of the selected models
as function of ${\displaystyle \frac{R}{R_{c}}}$  for suitable
values of $\{p,n,q,s,\lambda\}$ .

Let us now estimate $m_\sigma$ for the models discussed above. For
Model I \cite{Hu}, when the curvature is large, we find \be
\label{JGRG9} f_{I}(R)\sim - \frac{m^2 c_1}{c_2} + \frac{m^{2+2n}
c_1}{c_2^2 R^n} + \cdots \ , \ee and obtain the following
expression: \be \label{JGRG25} m_\sigma^2 \sim \frac{m^2 c_2^2}{2
n(n+1) c_1}\left(\frac{R}{m^2}\right)^{n+2}\ . \ee Here the order
of the mass-dimensional parameter $m^2$ should be $m^2\sim
10^{-64}\,{\rm eV}^2$. Then in Solar System, where $R\sim
10^{-61}\,{\rm eV}^2$, the mass is given by $m_\sigma^2 \sim
10^{-58 + 3n}\,{\rm eV}^2$ while on  the Earth atmosphere, where
$R \sim 10^{-50}\,{\rm eV}^2$, it has to be $m_\sigma^2 \sim
10^{-36 + 14n}\,{\rm eV}^2$. The order of the radius of the Earth
is $10^7\,{\rm m} \sim \left(10^{-14}\,{\rm eV}\right)^{-1}$.
Therefore the scalar field $\sigma$ is enough heavy  if $n\gg 1$
and the correction to the Newton law is not observed, being
extremely small. In fact, if we choose $n=10$, the order of the
Compton length of the scalar field $\sigma$ becomes that of the
Earth radius. On the other hand, in the Earth atmosphere, if we
choose $n=10$, for example, we find that the mass is extremely
large: \be \label{JGRG26} m_\sigma \sim 10^{43}\,{\rm GeV} \sim
10^{29} \times M_{\rm Planck}\ . \ee Here $M_{\rm Planck}$ is the
Planck mass. Hence, the Newton law correction should be extremely
small.

In   Model II \be \label{S1} f_{II}(R)= - \lambda R_0 \left[ 1 -
\left( 1 + \frac{R^2}{R_0^2}\right)^{-p} \right]\ , \ee if $R$ is
large compared with $R_0$, whose order of magnitude is that of the
curvature in the present universe, we find \be \label{S2}
f_{II}(R)=- \lambda R_0 + \lambda \frac{R_0^{2p+1}}{R^{2p}} +
\cdots \ . \ee By comparing Eq.(\ref{S2}) with Eq. (\ref{JGRG9}),
if the curvature is large enough when compared with $R_0$ or
$m^2$, as in the Solar System or on the Earth, we  can set  the
following identifications: \be \label{S3} \lambda R_0
\leftrightarrow \frac{m^2 c_1}{c_2} \ ,\quad \lambda R_0^{2p+1}
\leftrightarrow \frac{m^{2+2n} c_1}{c_2^2} \ ,\quad 2p
\leftrightarrow n\ . \ee We have $41 m^2 \sim R_0$. Then, if $p$
is large enough, there is no correction to the Newton law as in
Model I given by Eq.(\ref{eq:HS1}).

Let us now discuss  the instability of fluid matter proposed in
\cite{DK}, which may appear if the matter-energy density (or the
scalar curvature) is large enough when compared with the average
density the Universe, as it is inside the Earth. Considering the
trace of the above field equations and with a little algebra, one
obtains \be \label{JGRG27} \Box R +
\frac{F^{(3)}(R)}{F^{(2)}(R)}\nabla_\rho R \nabla^\rho R +
\frac{F'(R) R}{3F^{(2)}(R)} - \frac{2F(R)}{3 F^{(2)}(R)} =
\frac{\kappa^2}{6F^{(2)}(R)}T\,. \ee Here $T$ is the trace of the
matter energy-momentum tensor: $T\equiv T^{(m)\rho}_{ \rho}$. We
also denote the derivative $d^nF(R)/dR^n$ by $F^{(n)}(R)$. Let us
now consider the perturbation of the Einstein gravity solutions.
We denote the scalar curvature,  given by the matter density in
the Einstein gravity, by $R_b\sim (\kappa^2/2)\rho>0$ and separate
the scalar curvature $R$ into the sum of $R_b$ (background) and
the perturbed part $R_p$ as $R=R_b + R_p$
$\left(\left|R_p\right|\ll \left|R_b\right|\right)$. Then
Eq.(\ref{JGRG27}) leads to the perturbed equation: \bea
\label{JGRG28} 0 &=& \Box R_b +
\frac{F^{(3)}(R_b)}{F^{(2)}(R_b)}\nabla_\rho R_b \nabla^\rho R_b +
\frac{F'(R_b) R_b}{3F^{(2)}(R_b)} \nn && - \frac{2F(R_b)}{3
F^{(2)}(R_b)} - \frac{R_b}{3F^{(2)}(R_b)} + \Box R_p +
2\frac{F^{(3)}(R_b)}{F^{(2)}(R_b)}\nabla_\rho R_b \nabla^\rho R_p
+ U(R_b) R_p\ . \eea Here the potential $U(R_b)$ is given by \bea
\label{JGRG29} U(R_b) &\equiv&
\left(\frac{F^{(4)}(R_b)}{F^{(2)}(R_b)} -
\frac{F^{(3)}(R_b)^2}{F^{(2)}(R_b)^2}\right) \nabla_\rho R_b
\nabla^\rho R_b + \frac{R_b}{3} \nn && - \frac{F^{(1)}(R_b)
F^{(3)}(R_b) R_b}{3 F^{(2)}(R_b)^2} -
\frac{F^{(1)}(R_b)}{3F^{(2)}(R_b)} + \frac{2 F(R_b)
F^{(3)}(R_b)}{3 F^{(2)}(R_b)^2} - \frac{F^{(3)}(R_b) R_b}{3
F^{(2)}(R_b)^2} \eea It is convenient to consider the case where
$R_b$ and $R_p$ are uniform and do not depend on the spatial
coordinates. Hence, the d'Alembert operator can be replaced by the
second derivative with respect to the time, that is: $\Box R_p \to
-
\partial_t^2 R_p$.  Eq.(\ref{JGRG29}) assumes the following
structure: \be \label{JGRG30} 0=-\partial_t^2 R_p + U(R_b) R_p +
{\rm const}\ . \ee If $U(R_b)>0$, $R_p$ becomes exponentially
large with time, i.e. $R_p\sim \e^{\sqrt{U(R_b)} t}$, and the
system becomes unstable.

In the $1/R$-model, considering the background values, we find
\bea \label{JGRG32} && U(R_b) = - R_b + \frac{R_b^3}{6\mu^4} \sim
\frac{R_0^3}{\mu^4} \sim \left(10^{-26} \mbox{sec}\right)^{-2}
\left(\frac{\rho_m}{\mbox{g\,cm}^{-3}}\right)^3\ ,\nn && R_b \sim
\left(10^3 \mbox{sec}\right)^{-2}
\left(\frac{\rho_m}{\mbox{g\,cm}^{-3}}\right)\,. \eea Here the
mass parameter $\mu$ is of the order \be \label{JGRG31}
\mu^{-1}\sim 10^{18} \mbox{sec} \sim \left( 10^{-33} \mbox{eV}
\right)^{-1}\ . \ee Eq.(\ref{JGRG32}) tells us that the model is
unstable and it would decay in $10^{-26}$ sec (considering the
Earth size). In  Model I, however, $U(R_b)$ is negative: \be
\label{JGRG34} U(R_0) \sim - \frac{(n+2)m^2 c_2^2}{c_1 n(n+1)} < 0
\ . \ee Therefore, there is no matter instability.

For  Model (\ref{S1}), as it is clear from the identifications
(\ref{S3}), there is no matter instability too.

In order to study the stability of the de Sitter solution, let us
proceed as follows.  From the field equations (\ref{eq:motion}),
we obtain the trace \be \label{dS1} \Box f'(R) = \frac{1}{3}\left[
R - f'(R) R + 2 f(R) + \kappa^2 T\right]\,. \ee Here, as above,
$F(R)$ is $F(R) = R + f(R)$ and $T\equiv g^{\mu\nu}
T^{(m)}_{\mu\nu}$.

Now we consider the (in)stability around the de Sitter solution,
where $R=R_0$, and therefore $f(R_0)$ and $f'(R_0)$, are
constants. Then since the l.h.s. in Eq.(\ref{dS1}) vanishes for
$R=R_0$, we find \be \label{dS2} R_0 - f'(R_0) R_0 + 2 f(R_0) +
\kappa^2 T_0 = 0\ . \ee Let us expand  both sides of (\ref{dS2})
around $R=R_0$ as \be \label{dS3} R=R_0 + \delta R\,. \ee One
obtains \be \label{dS4} f''(R_0) \Box \delta R = \frac{1}{3}\left(
1 - f''(R_0) R_0 + f'(R_0) \right)\delta R\ . \ee Since \be
\label{dS5} \Box \delta R  = - \frac{d^2 \delta R}{dt^2} - 3H_0
\frac{d \delta R}{dt}\ , \ee in the de Sitter background, if \be
\label{dS6} C(R_0) \equiv \lim_{R\to R_0} \frac{1 - f''(R) R +
f'(R)}{f''(R)} > 0\ , \ee the de Sitter background is stable but,
if $C(R_0)<0$, the de Sitter background is unstable. The
expression for $C(R_0)$ could be valid even if $f''(R_0)=0$. More
precisely, the solution of (\ref{dS4}) is given by \be \label{dS7}
\delta R = A_+ \e^{\lambda_+ t} + A_- \e^{\lambda_- t}\ . \ee Here
$A_\pm$ are constants and \be \label{dS8} \lambda_\pm = \frac{-
3H_0 \pm \sqrt{9H_0^2 - C(R_0)}}{2}\ . \ee Then, if $C(R_0)<0$,
$\lambda_+$ is always positive and the perturbation grows up. This
leads to the instability. We have also to note that, when $C(R_0)$
is positive, if $C(R_0)>9H_0^2$, $\delta R$ oscillates and the
amplitude becomes exponentially small being: \be \label{dS9}
\delta R = \left(A \cos \omega_0 t + B \sin \omega_0
t\right)\e^{-3H_0 t/2}\,,\quad \omega \equiv \frac{\sqrt{C(R_0) -
9H_0^2}}{2}\ . \ee Here $A$ and $B$ are constant. On the other
hand, if $C(R_0)< 9H_0^2$, there is no oscillation in $\delta R$.

Let us now consider  the case where the matter contribution $T$
can be neglected in the de Sitter background and assume $f'(R)=0$
in the same background. We can assume that there are two de Sitter
background solutions satisfying $f'(R)=0$, for $R=R_1$ and $R=R_2$
as it could be the physical case if one asks for an inflationary
and a dark energy epoch. We also assume $f'(R)\neq 0$ if
$R_1<R<R_2$ or $R_2<R<R_1$. In the case $C(R_1)<0$ and $C(R_2)>0$,
the de Sitter solution, corresponding to $R=R_1$, is unstable but
the solution corresponding to $R=R_2$ is stable. Then there should
be a solution where the (nearly) de Sitter solution corresponding
to $R_1$ transits to the (nearly) de Sitter solution  $R_2$. Since
the solution corresponding to $R_2$ is stable, the universe
remains in the de Sitter solution corresponding to $R_2$ and there
is no more transition to any other de Sitter solution.

As an example, we consider Model I. For  large curvature values,
we find \be \label{dS10} f_{\rm I}(R)= - \Lambda +
\frac{\alpha}{R^{2n+1}}\ . \ee Here $\Lambda$ and $\alpha$ are
positive constants and $n$ is a positive integer. Then we find \be
\label{dS11} C(R) \sim \frac{1}{f''(R)} \sim
\frac{R^{2n+2}}{2n(2n+1) \alpha} > 0\,. \ee This means that  the
de Sitter solution in Model I can be stable. We have also to note
that $C(R_0) \sim H_0^{4n+4}/m^{4n+2}$. Here $m^2$ is the mass
scale introduced in \cite{Hu} and $m^2\ll H_0^2$: this means that
$C(R_0) \gg 9H_0^2$ and therefore there could  be no oscillation.

\begin{figure}[ht]
\includegraphics[scale=0.6]{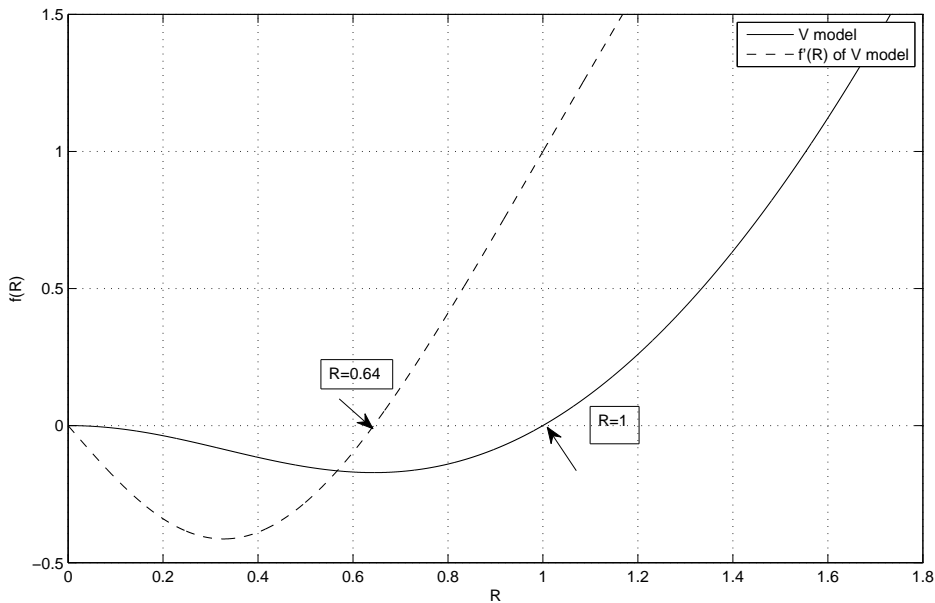}
\caption{Plots of Model V (\ref{JGRG53}) (solid line) and its
first derivative  (dashed line). Here $n=2$ and
$\alpha,\beta,\gamma$ are assumed as in (\ref{JGRG59}) with the
value of $R_0$ taken in the Solar System. $f'(R)$ is negative for
$0<R<0.64$. $f(R)$ is given in the range $0<R<1$ where we have
adopted suitable units.} \label{fig:odi}
\end{figure}

We may also consider the model proposed in
\cite{Nojiri:2007cq}(here Model V): \be \label{JGRG53} f_{V}(R)=
\frac{\alpha R^{2n} - \beta R^n}{1 + \gamma R^n} \, . \ee Here
$\alpha$, $\beta$, and $\gamma$ are positive constants and $n$ is
a positive integer. In Fig.\ref{fig:odi},  we  show the behavior
of Model V and of its  first derivative. When the curvature is
large ($R\to \infty$), $f(R)$ behaves as a power law. Since the
derivative of $f(R)$ is given by \be \label{JGRG55} f'_{V}(R) =
\frac{n R^{n-1}\left( \alpha \gamma R^{2n} - 2\alpha R^n - \beta
\right)} {\left(1+\gamma R^n \right)^2}\ , \ee we find that the
curvature $R_0$ in the present universe, which satisfies the
condition $f'(R_0)=0$, is given by \be \label{JGRG56} R_0=\left[
\frac{1}{\gamma}\left(1+ \sqrt{ 1 + \frac{\beta\gamma}{\alpha}
}\right)\right]^{1/n} \ , \ee and \be \label{JGRG57} f(R_0) \sim
-2 \tilde R_0 = \frac{\alpha}{\gamma^2}\left( 1 + \frac{\left(1 -
\beta\gamma/\alpha \right) \sqrt{ 1 + \beta\gamma/\alpha}}{2 +
\sqrt{ 1 + \beta\gamma/\alpha}} \right) \ . \ee As shown in
\cite{Nojiri:2007cq}, the magnitudes of the parameters is given by
\be \label{JGRG59} \alpha \sim 2 \tilde R_0 R_0^{-2n},\ \beta \sim
4 {\tilde R_0}^2 R_0^{-2n} R_I^{n-1},\ \gamma \sim 2 \tilde R_0
R_0^{-2n} R_I^{n-1} . \ee Here $R_I$ is the curvature in the
inflationary epoch and we have assumed $f(R_I) \sim
(\alpha/\gamma) R_I^n \sim R_I$.

$C(R_0)$ in (\ref{dS6}) is given by \be \label{dS12} C(R_0) \sim
\frac{1}{f''(R_0)} = \frac{1+\gamma R_0^n}{2n^2 \alpha R_0^{2n-2}
\left( \gamma R_0^n - 1 \right)}\,. \ee By using the relations
(\ref{JGRG59}), we find \be \label{dS13} C(R_0) \sim
\frac{R_0^2}{4n^2 \tilde R_0}\ , \ee which is positive and
therefore the de Sitter solution is stable. We  notice that
$C(R_0)<9H_0^2$ and therefore, there could occur oscillations as
in (\ref{dS9}).

Furthermore, we can take into account the following model
\cite{sebast} (Model VI): \be \label{tan1} f_{VI}(R)=-\alpha
\left[ \tanh \left(\frac{b\left(R-R_0\right)}{2}\right) + \tanh
\left(\frac{b R_0}{2}\right)\right] = - \alpha \left[
\frac{\e^{b\left(R-R_0\right)} - 1}{\e^{b\left(R-R_0\right)} + 1}
+ \frac{\e^{b R_0} - 1}{\e^{bR_0} + 1}\right] \ee where $\alpha$
and $b$ are positive constants. When $R\to 0$, we find that \be
\label{tan2} f_{VI}(R) \to - \frac{\alpha b R}{2\cosh^2
\left(\frac{b R_0}{2}\right) }\,, \ee and thus $f(0)=0$. On the
other hand, when $R\to +\infty$, \be \label{tan3} f_{VI}(R) \to -
2\Lambda_{\rm eff} \equiv -\alpha \left[ 1 + \tanh \left(\frac{b
R_0}{2}\right)\right]\ . \ee If $R\gg R_0$, in the present
universe, $\Lambda_{\rm eff}$ plays the role of the effective
cosmological constant. We also obtain \be \label{tan4} f'_{VI}(R)=
- \frac{\alpha b }{2\cosh^2 \left(\frac{b \left(R -
R_0\right)}{2}\right) }\,, \ee which has a minimum when $R=R_0$,
that is: \be \label{tan5} f'_{VI}(R_0)= - \frac{\alpha b}{2}\, .
\ee Then in order to avoid anti-gravity, we find \be \label{tan6}
0< 1 + f'_{VI}(R_0) = 1 - \frac{\alpha b}{2}\, . \ee

Beside the above model, we can consider a model which is able to
describe, in principle,  both the early inflation and the late
acceleration epochs. The following two-step model \cite{sebast}
(Model VII): \be \label{tan7} f_{VII}(R)=-\alpha_0 \left[\tanh
\left(\frac{b_0\left(R-R_0\right)}{2}\right) + \tanh
\left(\frac{b_0 R_0}{2}\right)\right]
 -\alpha_I \left[ \tanh \left(\frac{b_I\left(R-R_I\right)}{2}\right)
+ \tanh \left(\frac{b_I R_I}{2}\right)\right]\, , \ee could be
useful to this goal. Let us assume \be \label{tan8} R_I\gg R_0\
,\quad \alpha_I \gg \alpha_0\ ,\quad b_I \ll b_0\ , \ee and \be
\label{tan8b} b_I R_I \gg 1\ . \ee When $R\to 0$ or $R\ll R_0 \ll
R_I$, $f_{VII}(R)$ behaves as \be \label{tan9} f_{VII}(R) \to -
\left[\frac{\alpha_0 b_0 }{2\cosh^2 \left(\frac{b_0 R_0}{2}\right)
} + \frac{\alpha_I b_I }{2\cosh^2 \left(\frac{b_I R_I}{2}\right)
}\right]R\ \,, \ee and we find again $f_{VII}(0)=0$. When $R\gg
R_I$, we find \be \label{tan10} f(R)_{VII} \to - 2\Lambda_I \equiv
 -\alpha_0 \left[ 1 + \tanh \left(\frac{b_0 R_0}{2}\right)\right]
 -\alpha_I \left[ 1 + \tanh \left(\frac{b_I R_I}{2}\right)\right]
\sim -\alpha_I \left[ 1 + \tanh \left(\frac{b_I
R_I}{2}\right)\right]\ . \ee On the other hand, when $R_0\ll R \ll
R_I$, we find \be \label{tan11} f_{VII}(R) \to -\alpha_0 \left[ 1
+ \tanh \left(\frac{b_0 R_0}{2}\right)\right]
 - \frac{\alpha_I b_I R}{2\cosh^2 \left(\frac{b_I R_I}{2}\right) }
\sim -2\Lambda_0 \equiv -\alpha_0 \left[ 1 + \tanh \left(\frac{b_0
R_0}{2}\right)\right] \ . \ee Here, we have assumed the condition
(\ref{tan8b}). We also find \be \label{tan12} f'_{VII}(R)= -
\frac{\alpha_0 b_0 }{2\cosh^2 \left(\frac{b_0 \left(R -
R_0\right)}{2}\right) } - \frac{\alpha_I b_I }{2\cosh^2
\left(\frac{b_I \left(R - R_I\right)}{2}\right) }\, , \ee which
has two minima for $R\sim R_0$ and $R\sim R_I$. When $R= R_0$, we
obtain \be \label{tan13} f'_{VII}(R_0)= - \alpha_0 b_0 -
\frac{\alpha_I b_I }{2\cosh^2 \left(\frac{b_I \left(R_0 -
R_I\right)}{2}\right) }
> - \alpha_I b_I - \alpha_0 b_0 \ .
\ee On the other hand, when $R=R_I$, we get \be \label{tan14}
f'_{VII}(R_I)= - \alpha_I b_I - \frac{\alpha_0 b_0 }{2\cosh^2
\left(\frac{b_0 \left(R_0 - R_I\right)}{2}\right) }
> - \alpha_I b_I - \alpha_0 b_0 \ .
\ee Then, in order to avoid the anti-gravity behavior, we find \be
\label{tan15} \alpha_I b_I + \alpha_0 b_0 < 1\ . \ee

Let us now investigate the correction to the Newton potential and
the matter instability issue related to Models VI and VII. In the
Solar System domain, on or inside the Earth, where $R\gg R_0$,
$f(R)$ in Eq.(\ref{tan1}) can be approximated by \be \label{tan16}
f_{VI}(R) \sim -2 \Lambda_{\rm eff} + 2\alpha \e^{-b(R-R_0)}\, .
\ee On the other hand, since $R_0\ll R \ll R_I$, by assuming
Eq.~(\ref{tan8b}), $f(R)$ in (\ref{tan7}) can be also approximated
by \be \label{tan17} f_{VII}(R) \sim -2 \Lambda_0 + 2\alpha
\e^{-b_0(R-R_0)}\, , \ee which has the same expression, after
having identified $\Lambda_0 = \Lambda_{\rm eff}$ and $b_0=b$.
Then, we may check the case of (\ref{tan16}) only. In this case,
the effective mass has the following form \be \label{tan18}
m_\sigma^2 \sim \frac{\e^{b(R-R_0)}}{4\alpha b^2}\,, \ee which
could be again very large. In fact, in the Solar System, we find
$R\sim 10^{-61}\,{\rm eV}^2$. Even if we choose $\alpha\sim 1/b
\sim R_0 \sim \left(10^{-33}\,{\rm eV}\right)^2$, we find that
$m_\sigma^2 \sim 10^{1,000}\,{\rm eV}^2$, which is, ultimately,
extremely heavy. Then, there will  be no appreciable correction to
the Newton law. In the  Earth atmosphere, $R \sim 10^{-50}\,{\rm
eV}^2$, and even if we choose $\alpha\sim 1/b \sim R_0 \sim
\left(10^{-33}\,{\rm eV}\right)^2$ again, we find that $m_\sigma^2
\sim 10^{10,000,000,000}\,{\rm eV}^2$. Then, a correction to the
Newton law is never observed in such  models. In this case, we
find that the effective potential $U(R_b)$ has the form \be
\label{tan19} U(R_e) = - \frac{1}{2\alpha b}\left(2\Lambda +
\frac{1}{b}\right)\e^{-b(R_e-R_0)}\ , \ee which could be negative,
what would suppress any instability.

In order that a de Sitter solution exists in  $f(R)$-gravity, the
following condition has to be satisfied: \be \label{tan20}
R=Rf'(R) - 2f(R)\, . \ee For the model (\ref{tan1}), the r.h.s of
(\ref{tan20}) has the following form: \be \label{tan21}
R=-\frac{b\alpha
R}{2\cosh^2\left(\frac{b\left(R-R_0\right)}{2}\right)} + 2\alpha
\left[\tanh\left(\frac{b\left(R-R_0\right)}{2}\right) +\tanh
\left(\frac{bR_0}{2}\right)\right]\ . \ee For large $R$, the
r.h.s. behaves as \be \label{tan22} -\frac{b\alpha
R}{2\cosh^2\left(\frac{b\left(R-R_0\right)}{2}\right)} + 2\alpha
\left[\tanh\left(\frac{b\left(R-R_0\right)}{2}\right) +\tanh
\left(\frac{bR_0}{2}\right)\right] \to 2\alpha\ , \ee although the
l.h.s. goes to infinity. On the other hand, when $R$ is small, the
r.h.s. behaves as \be \label{tan23} -\frac{b\alpha
R}{2\cosh^2\left(\frac{b\left(R-R_0\right)}{2}\right)} + 2\alpha
\left[\tanh\left(\frac{b\left(R-R_0\right)}{2}\right) +\tanh
\left(\frac{bR_0}{2}\right)\right] \to \frac{b\alpha
R}{2\cosh^2\left(\frac{bR_0}{2}\right)}\ . \ee Then if \be
\label{tan24} \frac{b\alpha}{2\cosh^2\left(\frac{bR_0}{2}\right)}>
1 \ , \ee there is a de Sitter solution. Combining
Eq.(\ref{tan24}) with Eq.(\ref{tan6}), we find \be \label{tan25}
2>\alpha b > \frac{1}{2\cosh^2\left(\frac{bR_0}{2}\right)}\, . \ee
The stability, as above, is given by $C(R_{\rm dS})$, where
$R_{\rm dS}$ is the solution of (\ref{tan21}). The expression is
given by \be \label{tan26} C(R_{\rm dS}) = - R_{\rm dS} +
\frac{2\cosh^3 \left(\frac{b\left(R_{\rm dS}
-R_0\right)}{2}\right)}{ \alpha b^2 \sinh
\left(\frac{b\left(R_{\rm dS} -R_0\right)}{2}\right)}
 - \frac{1}{b\tanh\left(\frac{b\left(R_{\rm dS} -R_0\right)}{2}\right)}\ .
\ee Let us now rewrite Eq.(\ref{tan21}) as follows, \be
\label{tan27} R_{\rm dS} = 2\alpha \left[ \tanh
\left(\frac{b\left(R_{\rm dS} -R_0\right)}{2}\right) +
\tanh\left(\frac{bR_0}{2}\right) \right] \left[ 1 + \frac{\alpha
b}{2 \cosh^2 \left(\frac{b\left(R_{\rm dS}
-R_0\right)}{2}\right)}\right]^{-1}\ . \ee Then by using
(\ref{tan27}), we may rewrite (\ref{tan26}) in the following form:
\be \label{tan28} C(R_{\rm dS}) = \frac{ - \alpha^2 b^2
\left(1-x^2\right) \left[ \left(x - x_0\right)^2 + 1 -
x_0^2\right] + 4}{\alpha b^2 x \left(1-x^2\right)\left[2 + \alpha
b \left( 1 - x^2 \right) \right]}\ , \ee where \be \label{tan29} x
= \tanh \left(\frac{b\left(R_{\rm dS} -R_0\right)}{2}\right)\
,\quad x_0 = - \tanh \left(\frac{b R_0}{2}\right)\ , \ee and
therefore we have \be \label{tan30} -1 < x_0 \leq x < 1\ ,\quad
x_0 < 0\ . \ee

Let us now consider  (\ref{tan21}) in order to find a de Sitter
solution. Since  Eq.(\ref{tan21}) is difficult to solve in
general, we  assume $0<R_{\rm dS}\ll R_0$. Then we find \be
\label{tan31} R_{\rm dS}= \frac{\epsilon}{bx_0}\ ,\quad \epsilon
\equiv 1 - \frac{2\cosh^2\left(\frac{b R_0}{2}\right)}{\alpha b} =
1 - \frac{2}{\alpha b \left(1 - x_0^2\right)}\, . \ee
Eq.(\ref{tan24}) tells that the parameter $\epsilon$ is  positive
and, by assumption, very small: $0<\epsilon\ll 1$. Since
$\epsilon$ is small, by using Eqs.(\ref{tan29}), we find \be
\label{tan32} x=x_0 + \frac{\left(1-x_0^2\right)}{2x_0}\epsilon +
{\cal O}\left(\epsilon^2\right)\ . \ee Then by using the
expression  (\ref{tan28}) for $C(R_{\rm dS})$, we find \be
\label{tan33} C(R_{\rm dS}) \sim \frac{ - \alpha^2 b^2
\left(1-x_0^2\right)^2 + 4} {\alpha b^2 x_0
\left(1-x_0^2\right)\left[2 + \alpha b \left( 1 - x_0^2 \right)
\right]}\, . \ee From the definition of $\epsilon$ in
(\ref{tan31}), we find \be \label{tan34} \alpha b \left( 1 - x_0^2
\right) = 2 + 2\epsilon + {\cal O}\left(\epsilon^2\right)\,, \ee
and then, from Eq.(\ref{tan34}),  Eq.(\ref{tan33}) can be written
as follows; \be \label{tan35} C(R_{\rm dS}) \sim -
\frac{\epsilon}{bx_0}\, . \ee Since $x_0<0$ in the condition
(\ref{tan30}), we find $C(R_{\rm dS}) >0$ and therefore the de
Sitter solution is stable.

\begin{figure}[ht]
\includegraphics[scale=0.6]{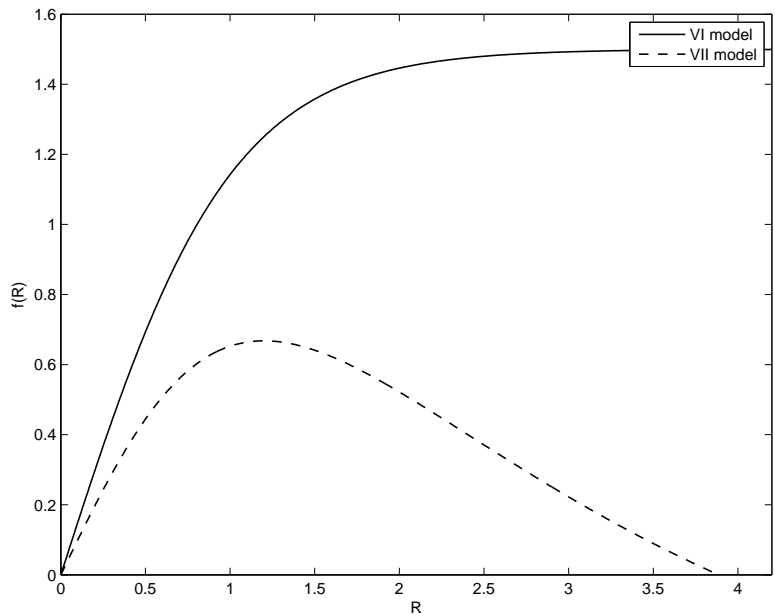}
\caption{Plots of Model VI (\ref{tan1}) (solid line) and Model VII
 (\ref{tan7}) (dashed line). Here $b=2$ and $b_{I}=0.5$ with
$\alpha=1.5$ and $\alpha_{I}=2$. The value of $R_I$ is taken in
the Solar System while $R_0$ corresponds to the present
cosmological value.} \label{fig:modelli67}
\end{figure}

In Fig. \ref{fig:modelli67}, we have plotted the two  models
(\ref{tan1}) and (\ref{tan7}) written in the form $F(R)=R+f(R)$.
We have used the inequalities (\ref{tan8})   assuming,  $R_I\sim
\rho_g \sim 10^{-24}$\,g/cm$^3$ for the  Galactic density in the
Solar vicinity and $R_0\sim \rho_g \sim 10^{-29}$\,g/cm$^3$ for
the present cosmological density. .

Our task is now to find reliable experimental bounds for such
models working at small and large scales. To this goal, we shall
take into account constraints coming from Solar System experiments
(which, at present, are capable of giving upper limits on the PPN
parameters) and constraints coming from interferometers, in
particular those giving limits on the (eventual) scalar components
of GWs. If constraints (and in particular the ranges of model
parameters given by them) are comparable, this could constitute,
besides  other experimental and observational probes,  a good hint
to achieve a self-consistent $f(R)$ theory at very different
scales.

\section{Constraining $f(R)$-models by PPN parameters}
The above models can be constrained at Solar System level by
considering the PPN formalism. This approach is extremely
important in order to test gravitational theories and to compare
them with GR. As it is shown in \cite{ppn-tot1,mpla}, one can
derive the PPN-parameters $\gamma$ and $\beta$ in terms of a
generic analytic function $F(R)$  and its derivative

\begin{equation}
\gamma-1=-\frac{F''(R)^{2}}{F'(R)+2F''(R)^{2}}\,,\label{eq:PPNgamma}\end{equation}

\begin{equation}
\beta-1=\frac{1}{4}\left[\frac{F'(R)\cdot
F''(R)}{2F'(R)+3F''(R)^{2}}\right]\frac{d\gamma}{dR}\,.\label{eq:PPNbeta}\end{equation}
These quantities have to fulfill the constraints  coming from the
Solar System experimental tests  summarized in Table I. They are
the perihelion shift of Mercury \cite{Shapiro}, the Lunar Laser
Ranging \cite{Williams}, the upper limits coming from the Very
Long Baseline Interferometry (VLBI) \cite{Shapiro S} and the
results obtained from the Cassini spacecraft mission in the delay
of the radio waves transmission near the Solar conjunction
\cite{Bertotti}.
\[
\]
\begin{table}
\begin{center}
\begin{tabular}{|c|c|}
\hline Mercury perihelion Shift&
$\left|2\gamma-\beta-1\right|<3\times10^{-3}$\tabularnewline
\hline \hline Lunar Laser Ranging &
$4\beta-\gamma-3=(0.7\pm1)\times10^{-3}$\tabularnewline \hline
Very Long Baseline Interferometer&
$\left|\gamma-1\right|<4\times10^{-4}$\tabularnewline \hline
Cassini Spacecraft&
$\gamma-1=(2.1\pm2.3)\times10^{-5}$\tabularnewline \hline
\end{tabular}
\end{center}
\caption{Solar System experimental constraints on the PPN
parameters.} \label{1}
\end{table}
\[
\]

Let us take into account before the  $f(R)$-models
(\ref{eq:HS1})-(\ref{eq:TS}). Specifically, we want to investigate
the values or the ranges of parameters in which they match the
Solar-System experimental constraints in Table \ref{1}. In other
words, we use these models to search  under what circumstances it
is possible to significantly address cosmological observations by
$f(R)$-gravity and, simultaneously, evade the local tests of
gravity.

By integrating Eqs. (\ref{eq:PPNgamma})-(\ref{eq:PPNbeta}), one
obtains $f(R)$ solutions depending on $\beta$ and $\gamma$ which
has to be confronted with $\beta_{exp}$ and $\gamma_{exp}$
\cite{mpla}. If we plug into such equations the models
(\ref{eq:HS1})-(\ref{eq:TS}) and the experimental values of PPN
parameters, we will obtain algebraic constraints for the
phenomenological parameters $\{n,p,q,\lambda, s\}$. This is the
issue which we want to take into account in this section.

From Eq.(\ref{eq:PPNgamma}), assuming $F'(R)+2F''(R)^{2}\neq0$ and
defining ${\displaystyle
A=\left|\frac{1-\gamma}{2\gamma-1}\right|}$, we obtain

\begin{equation}
\left[F''(R)\right]^{2}-AF'(R)=0\,.\label{eq:F20}
\end{equation}
The general solution of such an equation is a polynomial function
\cite{mpla}.

Considering  Model II given by (\ref{eq:STAROBINSKY}), we obtain

\begin{equation}
\left[1-\frac{2 p R \left(\frac{R^2}{R_{c}^2}+1\right)^{-p-1}
\lambda
   }{R_{c}}\right] \left|\frac{\gamma -1}{2 \gamma -1}\right|-\frac{4 p^2
   \left(\frac{R^2}{R_{c}^2}+1\right)^{-2 p} R_{c}^2 \left(R_{c}^2-(2 p+1)
   R^2\right)^2 \lambda ^2}{\left(R^2+R_{c}^2\right)^4}=0\,.\label{eq:gammaSTARO}
   \end{equation}
Our issue is now  to find the values of $\lambda$, $p$, and
$R/R_{c}$ for which the Solar System experimental constraints are
satisfied. Some preliminary considerations are in order at this
point. Considering the de Sitter solution achieved from
(\ref{eq:STAROBINSKY}), we have $R=const=R_{1}=x_{1}R_{c}$, and
$x_{1}>0$. It is straightforward to obtain

\begin{equation}
\lambda=\frac{x_{1}\left(1+x_{1}^{2}\right)^{p+1}}{2\left[\left(1+x_{1}^{2}\right)^{p+1}-
1-\left(p+1\right)x_{1}^{2}\right]}\,.\label{eq:lambdaSTA}\end{equation}
On the other hand, the stability conditions  $F_{,R}>0$ and
$F_{,RR}>0$ give the inequality

\begin{equation}
\left(1+x_{1}^{2}\right)^{p+2}>1+
\left(p+2\right)x_{1}^{2}+\left(p+1\right)\left(2p+1\right)x_{1}^{4}\,,\label{eq:lam2STA}\end{equation}
which has to be satisfied. In particular, for  $p=1$, it is
$x_{1}>\sqrt{3}$ and then ${\displaystyle
\lambda>\frac{8}{3\sqrt{3}}=1.5396}$. In addition, the value of
$x_{1}$ satisfying the relation (\ref{eq:lam2STA}) is also the
point where $\lambda(x_{1})$, in Eq.(\ref{eq:lambdaSTA}), reaches
its minimum.

To determine values of $R$ compatible with PPN constraints, let us
consider the trace of the field equations (\ref{eq:motion}) and
explicit solutions, given  the density profile $\rho(r)$, in the
Solar vicinity. One can set the boundary condition considering
$F_{,R_{\infty}}=F_{R_{g}}$

\begin{equation}
F_{,R_{g}}=F_{,R}(R=k^2\rho_{g})\end{equation} where  $\rho_g \sim
10^{-24}$\,g/cm$^3$ is the observed Galactic density in the Solar
neighborhoods. At this point, we can see when the relation
(\ref{eq:gammaSTARO}) satisfies the constraints for very Long
Baseline Interferometer ($\gamma-1=4\times10^{-4}$) and Cassini
Spacecraft ($\gamma-1=2.1\times10^{-}5$). This allows  to find out
suitable values for $p$.

An important remark is in order at this point. These constraint
equations work if stability conditions hold. In the range
\begin{equation}
0<\frac{R}{R_{c}}<\frac{1}{\sqrt{2p+1}}\label{eq:STAf2cond}\end{equation}
$F_{,RR}$ is negative for the model (\ref{eq:STAROBINSKY}) and
then stability conditions are violated. To avoid this range, we
need, at least, $\frac{R}{R_{c}}>1$. For example, we can choose
$\frac{R}{R_{c}}=3.38$, corresponding to de Sitter behavior. Then
we have $p=1$ and $\lambda=2$. On the other hand, for
$0.944<\lambda<0.966$, we have $p=2$ and
$\frac{R}{R_{c}}=\sqrt{3}$; finally, for $R>>R_{c}$, we have
$\lambda=2$ and $p=1.5$. For these values of parameters, the Solar
System tests are evaded.

\begin{figure}[ht]
\includegraphics[scale=0.7]{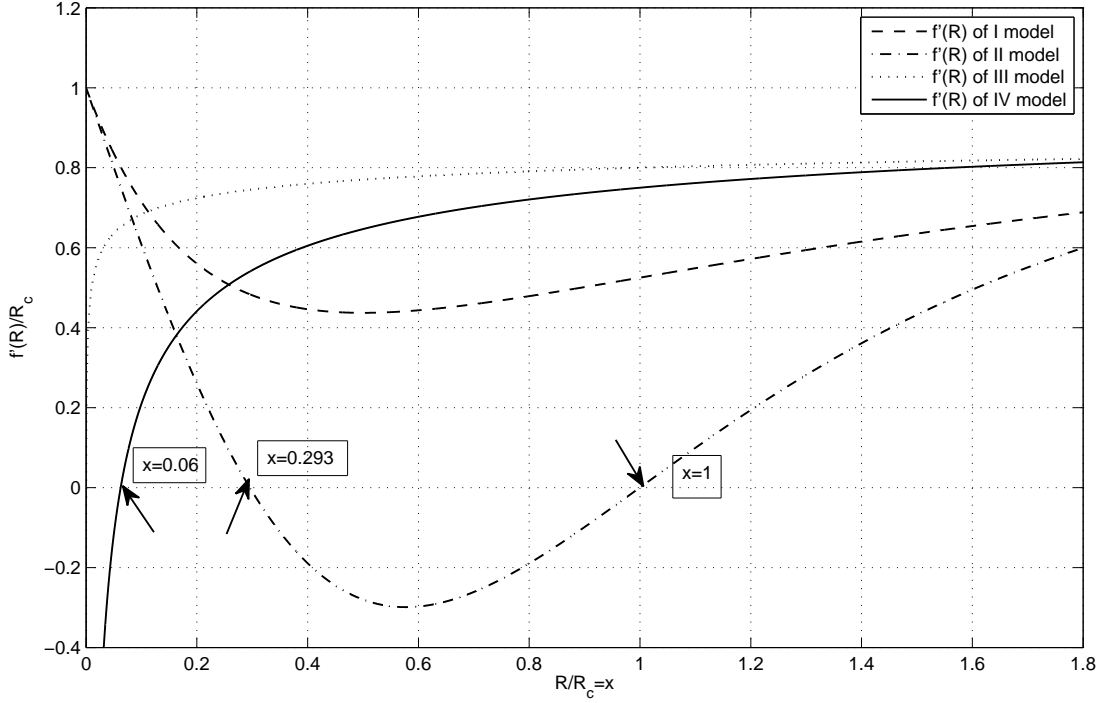}
\caption{Plots of the first derivatives  of  four different models
as function of $x=\frac{R}{R_{c}}$. Model I (dashed) is drawn for
$n=1$ and $\lambda=2$. Model II (dashdot),  for $p=2$,
$\lambda=0.95$. Model III (dotted), for   $s=0.5$ and
$\lambda=1.5$. Model IV (solid) is for $q=0.5$ and $\lambda=0.5$.
The labelled values of $x$ indicate where the derivative changes
its sign.} \label{fig:f(R)derivateprime}
\end{figure}

\begin{figure}[ht]
\includegraphics[scale=0.7]{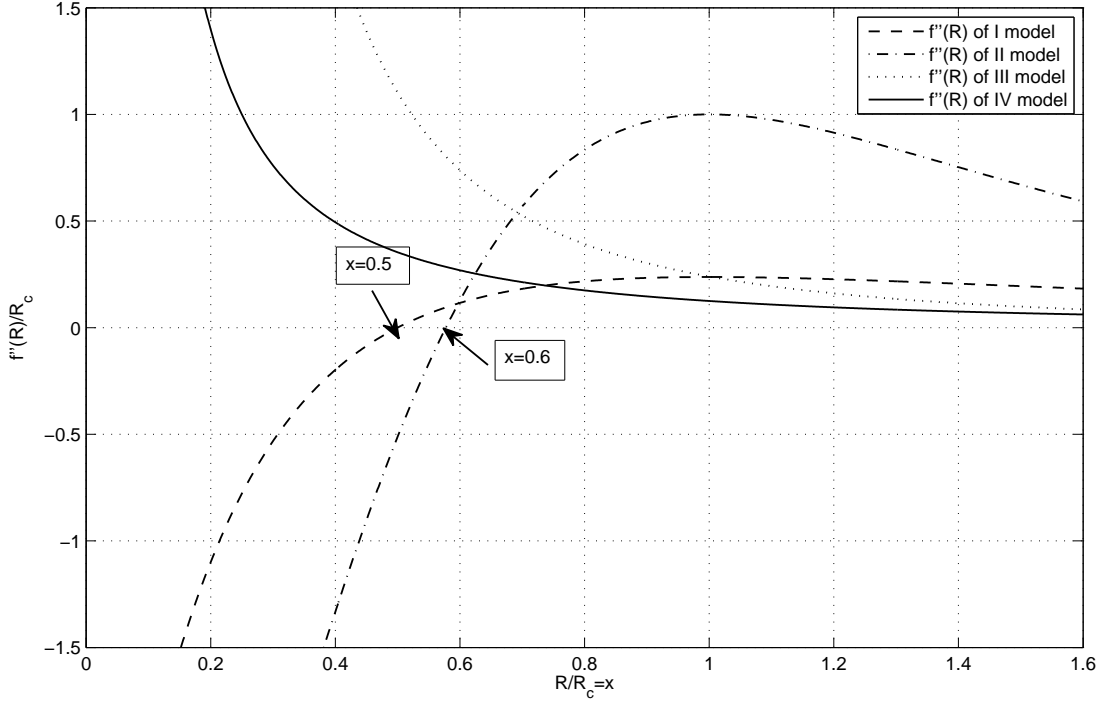}
\caption{As above for the second derivatives of the models.}
\label{fig:f(R)derivate2}
\end{figure}

Let us  consider now  Model I, given by (\ref{eq:HS}).  Inserting
it into the relation (\ref{eq:F20}), we get

\begin{equation}
 \frac{R^3 \left[\left(\frac{R}{R_{c}}\right)^{2 n}+1\right]^4 \left[R
   \left(\left(\frac{R}{R_{c}}\right)^{2 n}+1\right)^2-2 n
   \left(\frac{R}{R_{c}}\right)^{2 n}R_{c} \lambda \right] \left|\frac{\gamma -1}{2
   \gamma -1}\right|-4 n^2 \left[(2 n+1) \left(\frac{R}{R_{c}}\right)^{2 n}-2
   n+1\right]^2 \left(\frac{R}{R_{c}}\right)^{4 n}R_{c}^2 \lambda ^2}{R^4
   \left[\left(\frac{R}{R_{c}}\right)^{2 n}+1\right]^6}=0
   \end{equation}
Using the same procedure as above, $\lambda$ is related to the de
Sitter behavior. This means

\begin{equation}
\lambda=\frac{\left(1+x_{1}^{2n}\right)^{2}}{x_{1}^{2n-1}\left(2+2x_{1}^{2n}-2n\right)}
\,,\label{eq:lambdaHS}\end{equation}
while, from the stability
conditions, we get
\begin{equation}
 2x_{1}^{4}-\left(2n-1\right)\left(2n+4\right)x_{1}^{2n}+\left(2n-1\right)\left(2n-2\right)\geq 0\,.
 \label{eq:condHS}\end{equation}
For $n=1$, one obtains $x_{1}>\sqrt{3}$ ,
$\lambda>\frac{8}{3\sqrt{3}}$. In this model,  $F_{,RR}$ is
negative for
\begin{equation}
0<\frac{R}{R_{c}}<\left(\frac{2n-1}{2n+1}\right)^{\frac{1}{2n}}\,.\label{eq:HSf2cond}\end{equation}
The VLBI constraint  is satisfied for $n=1$ and $\lambda=2$,
while, for $n=1$ and $\lambda=1.5$, Cassini constraint holds.

By inserting   Model III, given by Eq.(\ref{eq:fR}), into the
relation (\ref{eq:F20}), we obtain
\begin{equation}
\label{general} \frac{R^3 \left[R-2 s R_{c} \left(\frac{R_{c}
}{R}\right)^{2 s}
   \lambda \right] \left|\frac{\gamma -1}{2 \gamma -1}\right|-4 \left(2
   s^2+s\right)^2 R_{c} ^2 \left(\frac{R_{c} }{r}\right)^{4 s}
   \lambda ^2}{R^4}=0\,.
   \end{equation}
The de-Sitter point  corresponds to
\begin{eqnarray}
\label{lamre}
\lambda=\frac{x_1^{2s+1}}{2(x_1^{2s}-s-1)}\,.
\end{eqnarray}
while the stability condition is $x_1^{2s}>2s^2+3s+1$.  VLBI and
Cassini constraints are satisfied by the sets of values: $s=1$,
$\lambda=1.53$, for $\frac{R}{R_{c}}\sim 1$; $s=2$,
$\lambda=0.95$, for  $\frac{R}{R_c}=\sqrt{3}$, ; $s=1$,
$\lambda=2$, for $\frac{R}{R_c}=3.38$.

Finally let us consider  Model VI, given by Eq.(\ref{tan1}), and
Model VII, given by Eq.(\ref{tan7}). Using Eq.(\ref{eq:F20}) for
(\ref{tan1}), we get

\begin{equation}
-\frac{1}{4} b \alpha  \text{sech}^2\left(\frac{1}{2} b
   (R-R_0)\right) \left[b^3 \alpha  \text{sech}^2\left(\frac{1}{2}
   b (R-R_0)\right) \tanh ^2\left(\frac{1}{2} b
   (R-R_0)\right)-2 \left|\frac{\gamma -1}{2 \gamma
   -1}\right|\right]=0\,.\label{PPN6}
  \end{equation}
As above, considering the stability conditions and the de Sitter
behavior, we  get the parameter ranges $0<b<2$ and $0<\alpha\leq
2$ which satisfy both VLBI and Cassini constraints. Inserting now
Model VII  in (\ref{eq:F20}), we have

\begin{equation}
\label{PPN7}
  \begin{split}
&\frac{1}{2} \left|\frac{\gamma -1}{2 \gamma -1}\right| \left[b
\alpha
   \text{sech}^2\left(\frac{1}{2} b (R-R_0)\right)-b_I
   \alpha _I \text{sech}^2\left(\frac{1}{2} b_I
   (R-R_I)\right)+2\right]\\
   &-\frac{1}{4} \left[b^2 \alpha
   \text{sech}^2\left(\frac{1}{2} b (R-R_0)\right) \tanh
   \left(\frac{1}{2} b (R-R_0)\right)-b_I^2\alpha _I
   \text{sech}^2\left(\frac{1}{2} b_I (R-R_I)\right) \tanh
   \left(\frac{1}{2} b_I (R-R_I)\right)\right]^2=0\,.
   \end{split}
   \end{equation}

\begin{figure}[ht]
\includegraphics[scale=0.7]{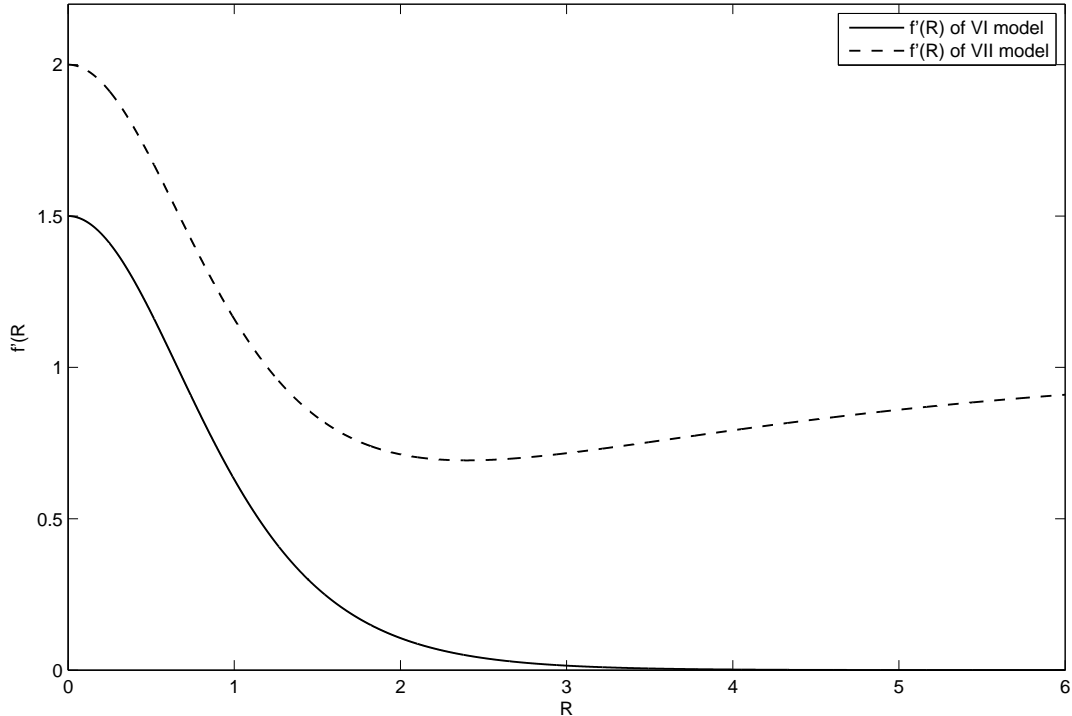}
\caption{Plots  represent the first derivatives  of functions
(\ref{tan6}) (solid line) and (\ref{tan7})  (dashed line). Here,
$b=2$, $b_{I}=0.5$, $\alpha=1.5$ and $\alpha_{I}=2$ with $R_I$
with the Solar System value and  $R_0$ the today cosmological
value. It is $F_,R>0$ for  $R>0$. } \label{fig:derivateprime67}
\end{figure}

\begin{figure}[ht]
\includegraphics[scale=0.7]{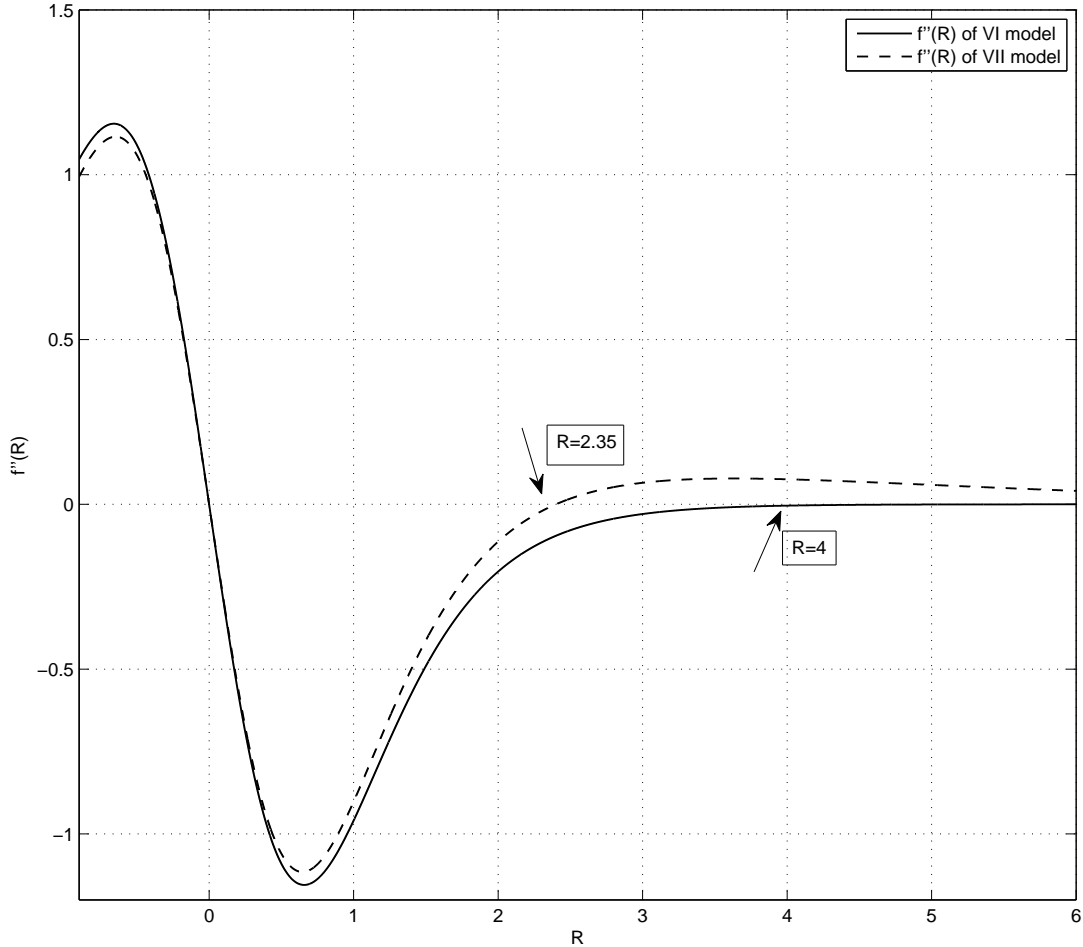}
\caption{Second derivatives of  Model VI (solid line) and VII
(dashed line). Here $F_{,RR}$ is negative  in the range $0<R<4$
for Model VI  and  in the range $0<R<2.35$ for  Model VII. As
above, we have used $b=2$, $b_{I}=0.5$, $\alpha=1.5$ and
$\alpha_{I}=2$ with the value of $R_I$ taken in the Solar System
and $R_0$ for the today  cosmological value.}
\label{fig:derivateseconde67}
\end{figure}

From the stability condition, we have that $F_{,R}>0$ for $R>0$,
(see Fig.\ref{fig:derivateprime67})  and $F_{,RR}<0$  for
$0<R<2.35$ in suitable units (see
Fig.\ref{fig:derivateseconde67}). Observational constraints  from
VLBI and Cassini experiments are fulfilled for \be R_I\gg R_0\
,\quad \alpha_I \gg \alpha\ ,\quad b_I \ll b\,.\ee Plots for
$b=2$, $b_I=0.5$, $\alpha=1.5$ and $\alpha_I=2$, verifying the
constraints, are reported in Figs. \ref{fig:derivateprime67} and
\ref{fig:derivateseconde67}.


Considering now the relation for $\beta$ given by Eq.
(\ref{eq:PPNbeta}), one can easily verify that it is

\begin{equation}
\frac{d\gamma}{dR}=-\frac{d}{dR}\left[\frac{F''(R)^{2}}{F'(R)+2F''(R)^{2}}\right]=0\,,\label{eq:Dgamma}\end{equation}

and this result  implies

\begin{equation}
4(\beta-1)=0\,.\label{eq:condbeta}\end{equation} This means the
complete compatibility of the $f(R)$ solutions between the
PPN-parameters $\beta$ and $\gamma$.

Now we want to see if the  parameter values, obtained for these
models, are compatible with bounds coming from the stochastic
background of GWs achieved by interferometric experiments.

\section{Stochastic  backgrounds of gravitational waves to constrain $f(R)$-gravity}
As we said before,  also the stochastic background of GWs can be
taken into account in order to constrain models. This approach
could reveal  very interesting because production of primordial
GWs could be a robust prediction for any model attempting to
describe the cosmological evolution  at primordial epochs.
However, bursts of gravitational radiation emitted from a large
number of unresolved and uncorrelated astrophysical sources
generate a stochastic background at more recent epochs,
immediately following the onset of galaxy formation. Thus,
astrophysical backgrounds might overwhelm the primordial one and
their investigation provides important constraints on the signal
detectability coming from the very early Universe, up to the
bounds of the Planck epoch and the initial singularity
\cite{Maggiore,Allen,Grishchuk,AO}.

It is worth stressing the unavoidable and fundamental character of
such a mechanism. It directly derives from the inflationary
scenario \cite{Watson,Guth}, which well fits the WMAP data with
particular good agreement with almost exponential inflation and
spectral index $\approx 1$, \cite{Bennet,Spergel}.

The main characteristics of the gravitational backgrounds produced
by cosmological sources depend both on the emission properties of
each single source and on the source rate evolution with redshift.
It is therefore interesting to compare and contrast the probing
power of these classes of  $f(R)$-models at  hight, intermediate
and zero redshift \cite{tuning}.

To this purpose,  let us  take into account the primordial
physical process which gave rise to a characteristic spectrum
$\Omega_{sgw}$ for the early stochastic background of relic scalar
GWs by which we can recast the further degrees of freedom coming
from fourth-order gravity. This approach can greatly contribute to
constrain viable cosmological models. The physical process related
to the production has been analyzed, for example, in
\cite{Allen,Grishchuk,Allen2} but only for the first two tensorial
components due to standard General Relativity. Actually the
process can be improved considering also the third scalar-tensor
component strictly related to the further $f(R)$ degrees of
freedom \cite{CCD}.

Before starting with the analysis, it has to be emphasized that
the stochastic background of scalar GWs can be described in terms
of a scalar field $\Phi$ and characterized by a dimensionless
spectrum (see the analogous definitions for tensorial waves in
\cite{Maggiore,Allen,Grishchuk,AO}). We can write the energy
density of scalar GWs  in terms of the closure energy density of
GWs per logarithmic frequency interval as
\begin{equation}
\Omega_{sgw}(f)=\frac{1}{\rho_{c}}\frac{d\rho_{sgw}}{d\ln
f}\,,\label{eq: spettro}\end{equation} where
 \begin{equation}
\rho_{c}\equiv\frac{3H_{0}^{2}}{8\pi G}\label{eq: densita'
critica}\end{equation} is the  critical energy density of the
Universe, $H_0$ the today observed Hubble expansion rate, and
$d\rho_{sgw}$ is the energy density of the gravitational radiation
scalar part  contained in the frequency range from $f$ to $f+df$.
We are considering now standard units.

The calculation for a simple inflationary model can be performed
assuming that the early Universe is described by an inflationary
de Sitter phase emerging in the radiation dominated era
\cite{Allen,Grishchuk,AO}. The conformal metric element is
\begin{equation}
ds^{2}=a^{2}(\eta)[-d\eta^{2}+d\overrightarrow{x}^{2}+
h_{\mu\nu}(\eta,\overrightarrow{x})dx^{\mu}dx^{\nu}], \label{eq:
metrica}\end{equation} and a  GW with  tensor and scalar modes in
the $z+$ direction   is given by \cite{CCD}
\begin{equation}
\tilde{h}_{\mu\nu}(t-z)=A^{+}(t-z)e_{\mu\nu}^{(+)}+A^{\times}(t-z)e_{\mu\nu}^{(\times)}+\Phi(t-z)e_{\mu\nu}^{(s)}\,.\label{eq:
perturbazione totale}\end{equation}
The pure scalar component is
then
\begin{equation}
h_{\mu\nu}=\Phi e_{\mu\nu}^{(s)}\,,\label{eq: perturbazione
scalare}\end{equation} where $e_{\mu\nu}^{(s)}$ is the
polarization tensor.

\begin{figure}[ht]
\includegraphics[scale=0.7]{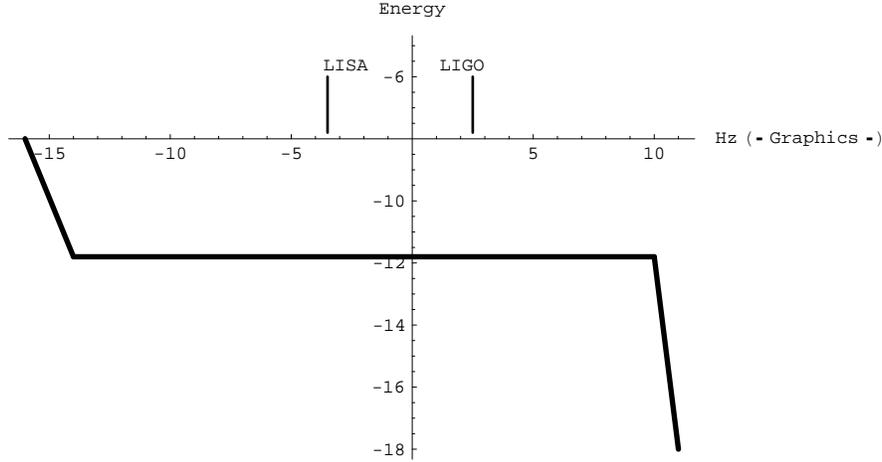}
\caption{The spectrum of relic scalar GWs in inflationary models
is flat over a wide range of frequencies. The horizontal axis is
$\log_{10}$ of frequency, in Hz. The vertical axis is
$\log_{10}\Omega_{gsw}$. The inflationary spectrum rises quickly
at low frequencies (wave which re-entered in the Hubble sphere
after the Universe became matter dominated) and falls off above
the (appropriately redshifted) frequency scale $f_{max}$
associated with the fastest characteristic time of the phase
transition at the end of inflation. The amplitude of the flat
region depends only on the energy density during the inflationary
stage; we have chosen the largest amplitude consistent with the
WMAP constrains on scalar perturbations. This means that, at LIGO
and LISA frequencies, we have $\Omega_{sgw}<2.3*10^{-12}$}
\label{fig:spectrum}
\end{figure}

It is possible to write an expression for the energy density of
the stochastic relic  scalar  gravitons  in the frequency interval
$(\omega,\omega+d\omega)$ as
\begin{equation}
d\rho_{sgw}=2\hbar\omega\left(\frac{\omega^{2}d\omega}{2\pi^{2}c^{3}}\right)N_{\omega}=
\frac{\hbar
H_{ds}^{2}H_{0}^{2}}{4\pi^{2}c^{3}}\frac{d\omega}{\omega}=\frac{\hbar
H_{ds}^{2}H_{0}^{2}}{4\pi^{2}c^{3}} \frac{df}{f}\,,\label{eq: de
energia}\end{equation} where $f$, as above, is the frequency in
the standard comoving time. Eq.(\ref{eq: de energia}) can be
written in terms of the today and de Sitter values of energy
density being
\begin{equation} H_{0}=\frac{8\pi G\rho_{c}}{3c^{2}}\,,\qquad H_{ds}=\frac{8\pi G\rho_{ds}}{3c^{2}}.\end{equation}
Introducing the Planck density ${\displaystyle \rho_{\rm
Planck}=\frac{c^{7}}{\hbar G^{2}}}$, the spectrum is given by
\begin{equation}
\Omega_{sgw}(f)=\frac{1}{\rho_{c}}\frac{d\rho_{sgw}}{d\ln
f}=\frac{f}{\rho_{c}}\frac{d\rho_{sgw}}{df}=\frac{16}{9}\frac{\rho_{ds}}{\rho_{\rm
Planck}}.\label{eq: spettro gravitoni}\end{equation} At this
point,  some  comments are in order. First of all, such a
calculation works for a simplified model which does not include
the matter dominated era. If  such an era is also included, the
redshift at equivalence epoch has to be considered. Taking into
account also results in \cite{Allen2}, we get
\begin{equation}
\Omega_{sgw}(f)=\frac{16}{9}\frac{\rho_{ds}}{\rho_{\rm
Planck}}(1+z_{eq})^{-1},\label{eq: spettro gravitoni
redshiftato}\end{equation} for the waves which, at the epoch in
which the Universe becomes matter dominated, have a frequency
higher than $H_{eq}$, the Hubble parameter at equivalence. This
situation corresponds to frequencies $f>(1+z_{eq})^{1/2}H_{0}$.
The redshift correction in Eq.(\ref{eq: spettro gravitoni
redshiftato}) is needed since the today observed Hubble parameter
$H_{0}$ would result different  without a matter dominated
contribution. At lower frequencies, the spectrum is given by
\cite{Allen,Grishchuk}
\begin{equation}
\Omega_{sgw}(f)\propto f^{-2}.\label{eq: spettro basse
frequenze}\end{equation}
Nevertheless, since the spectrum falls off $\propto f^{-2}$ at low
frequencies, this means that today, at LIGO-VIRGO and LISA
frequencies  (indicated in Fig. \ref{fig:spectrum}), one gets
\begin{equation} \Omega_{sgw}(f)h_{100}^{2}<2.3\times
10^{-12}.\label{eq: limite spettro WMAP}\end{equation} It is
interesting to calculate the  corresponding strain at $\approx
100Hz$, where interferometers like VIRGO and LIGO reach a maximum
in sensitivity (see e.g. \cite{Ligo1,Ligo2}). The well known
equation for the characteristic amplitude \cite{Allen,Grishchuk},
adapted to the scalar component of GWs, can be used. It is
\begin{equation}
\Phi_{c}(f)\simeq1.26\times
10^{-18}\left(\frac{1Hz}{f}\right)\sqrt{h_{100}^{2}\Omega_{sgw}(f)},\label{eq:
legame ampiezza-spettro}\end{equation} and then we obtain the
values in the Table \ref{2}.

\[
\]
\begin{table}
\begin{center}
\begin{tabular}{|c|c|}
\hline
$\Phi_{c}(100Hz)<2\times10^{-26}$&
LIGO\tabularnewline
\hline
$\Phi_{c}(100Hz)<2\times10^{-25}$&
VIRGO\tabularnewline
\hline
$\Phi_{c}(100Hz)<2\times10^{-21}$&
LISA\tabularnewline
\hline
\end{tabular}
\end{center}
\caption{Upper limits on the expected  amplitude for the GW scalar
component.} \label{2}
\end{table}
\[
\]

In summary, the above results point out that a further scalar
component of GWs, coming e.g. from $f(R)$-gravity, should be
seriously considered in the signal detection of interferometers.
As  discussed in \cite{tuning}, this fact could constitute either
an independent  test for alternative theories of gravity or a
further probe of GR capable of ruling out other theories.

At this point, using the above LIGO, VIRGO and LISA upper bounds,
calculated for the characteristic amplitude of GW scalar
component, let us test the  $f(R)$-gravity models, considered in
the previous sections,  to see whether they are compatible both
with the Solar System and GW stochastic background.

Before starting with the analysis, taking into account the
discussion in Sec.II,  we have that the GW scalar component is
derived considering
 \begin{equation}
\Phi=-\frac{\delta\sigma}{\sigma_{0}}\,, \label{eq:PHI} \qquad
\sigma=-\ln(1+f'(A))=\ln
F'(A)\,,\qquad\delta\sigma=\frac{f''(A)}{1+f'(A)}\delta A\,.
\end{equation}
As standard, we are assuming  small perturbations in the conformal
frame \cite{CCD}). This means
\begin{equation}
g_{\mu\nu}=\eta_{\mu\nu}+h_{\mu\nu}\,,\qquad\sigma=\sigma_{0}+\delta\sigma\,.\label{eq:
linearizza}\end{equation} These assumptions allow to derive the
"linearized" curvature invariants
$\widetilde{R}_{\mu\nu\rho\sigma}$ , $\widetilde{R}_{\mu\nu}$ and
$\widetilde{R}$  and then the linearized field equations
\cite{Misner}

\begin{equation}
\begin{array}{c}
\widetilde{R}_{\mu\nu}-\frac{\widetilde{R}}{2}\eta_{\mu\nu}=-\partial_{\mu}\partial_{\nu}\Phi+\eta_{\mu\nu}\square \Phi\\
\\{}\square \Phi=m^{2}\Phi\,.\end{array}\label{eq: linearizzate1}\end{equation}
As above, for the considered models,  we have to determine the
values of the characteristic parameters  which are compatible with
both Solar System and GW stochastic background.

Let us start, for example,  with the model (\ref{eq:fR}). Starting
from the definitions (\ref{eq:PHI}), it is straightforward to
derive the scalar component amplitude

\begin{equation}
\Phi_{III}=\frac{s (2 s+1) \left(\frac{R_{c}}{R}\right)^{2 s+1} \lambda
   }{\left[s R_{c} \left(\frac{R_{c}}{R}\right)^{2 s} \lambda
   -R\right] \log \left[2-2 s \left(\frac{R_{c}}{R}\right)^{2 s+1}
   \lambda \right]}\,.\label{eq:GWgeneral}
\end{equation}
Such an equation satisfies the  constraints in Table.\ref{2} for
the values $s=0.5$,  $\frac{R}{R_{c}}\sim1$, $\lambda=1.53$ and
$s=1$, $\frac{R}{R_{c}}\sim1$, $\lambda=0.95$ (LIGO); $s=2$,
$\frac{R}{R_c}=\sqrt{3}$, $\lambda=2$ (VIRGO); $s=1$, $\lambda=2$
and $\frac{R}{R_c}=3.38$ (LISA).

It is important to stress the nice agreement with the figures
achieved from the  PPN constraints. In this case, we have assumed
$R_{c}\sim \rho_c \sim 10^{-29}$\,g/cm$^3$, where $\rho_c$ is the
present day cosmological density.

Considering  the model (\ref{eq:HS}), we obtain
\begin{equation}
\Phi_{I}=-\frac{n \left[(2 n+1) \left(\frac{R}{R_{c}}\right)^{2
n}-2 n+1\right]
   \left(\frac{R}{R_{c}}\right)^{2 n-1} \lambda
   }{\left[\left(\frac{R}{R_{c}}\right)^{2 n}+1\right] \left\{R
   \left[\left(\frac{R}{R_{c}}\right)^{2 n}+1\right]^2-n
   \left(\frac{R}{R_{c}}\right)^{2 n} R_{c} \lambda \right\} \log \left(1-\frac{2 n
   \left(\frac{R}{R_{c}}\right)^{2 n-1} \lambda
   }{\left(\left(\frac{R}{R_{c}}\right)^{2 n}+1\right)^2}\right)}\,.\label{eq:GWHS}\end{equation}
The expected constraints for GW scalar amplitude are fulfilled for
$n=1$ and  $\lambda=2$ and for $n=1$ and $\lambda=1.5$ when $
0.3<\frac{R}{R_{c}}<1$.

Furthermore, considering the model (\ref{eq:STAROBINSKY}), one
gets
\begin{equation}
\Phi_{I}=-\frac{2p\left(1+\frac{R^{2}}{R_{c}^{2}}\right)^{-p}R_{c}\left(\left(1+2p\right)R^{2}-R_{c}^{2}\right)\lambda}
{\left(R^{2}-R_{c}^{2}\right)^{2}\left[2-\frac{2p\left(1+\frac{R^{2}}{R_{c}^{2}}\right)^{-1-p}\lambda}{R_{c}}\right]
\ln\left[2-\frac{2pR\left(1+\frac{R^{2}}{R_{c}^{2}}\right)^{-1-p}\lambda}{R_{c}}\right]}\,.\end{equation}
 The LIGO upper bound  is fulfilled for
 $p=1$, $\frac{R}{R_{c}}>\sqrt{3}$,
 $\lambda>\frac{8}{3\sqrt{3}}$; the VIRGO one for $p=1$,
$\frac{R}{R_{c}}=3.38$, $\lambda=2$; finally, for LISA, we have
$p=2$,  $\frac{R}{R_{c}}=\sqrt{3}$ and $0.944< \lambda< 0.966$.
Besides, considering LISA in the regime $R>>R_{c}$, we have
$\lambda=2$ and $p=1.5$.

Finally,  let us consider Models VI and VII. We have

\begin{equation}
\Phi_{VI}=\frac{b^2 \alpha  \tanh \left[\frac{1}{2} b
(R-R_0)\right]}{[b
   \alpha +\cosh (b (R-R_0))+1] \ln \left[\frac{b \alpha }{\cosh
   (b (R-R_0))+1}\right]}\,,\label{PHI6}
   \end{equation}
and

 \begin{equation}
 \label{PHI7}
 \begin{split}
\Phi_{VII}=& \log \left[0.5 \left(b \alpha  \text{sech}^2(0.5 b (R-R_0))-
b_I \alpha_I \text{sech}^2(0.5b_I (R-R_I))+2\right)\right] \\
&\times\left[b \alpha  \text{sech}^2(0.5 b (R-
   R_0))-b_I \alpha_I \text{sech}^2(0.5 b_I
   (R-R_I))+4\right] \\
   &\times\left[b^2 \alpha  \text{sech}^2(0.5 b
   (R-R_0)) \tanh (0.5 b (R-R_0))-b_I^{2}
   \alpha_I \text{sech}^2(0.5 b_I(R-R_I)) \tanh
   (0.5b_I (R-R_I))\right]\,.
   \end{split}
   \end{equation}

These equations  satisfy the constraints for  VIRGO, LIGO and LISA
for  $b=2$,  $b_{I}=0.5$, $\alpha=1.5$ and $\alpha_{I}=2$ with
$R_I$ valued at Solar System scale and  $R_0$ at cosmological
scale.

\section{Conclusions}

In this paper, we have investigated the possibility that some
viable $f(R)$ models could be constrained considering both Solar
System experiments and upper bounds on the stochastic background
of gravitational radiation. Such bounds   come from
interferometric ground-based (VIRGO and LIGO) and space (LISA)
experiments. The underlying philosophy is to show that the $f(R)$
approach, in order to describe consistently  the observed
universe, should  be tested at very different scales, that is at
very different redshifts. In other words, such a proposal could
partially contribute to remove the unpleasant degeneracy affecting
the wide class of dark energy models, today on the ground.

Beside the request to evade the Solar System tests, new methods
have been recently proposed to investigate the evolution and the
power spectrum of cosmological perturbations in $f(R)$ models
\cite{huspe}. The investigation of stochastic background, in
particular of the scalar component of GWs coming from the $f(R)$
additional degrees of freedom, could acquire, if revealed by the
running and forthcoming experiments,  a fundamental importance to
discriminate among the various  gravity theories \cite{tuning}.
These data (today only upper bounds coming from simulations) if
combined with Solar System tests, CMBR anisotropies, LSS, etc.
could greatly help to achieve a self-consistent cosmology
bypassing the shortcomings of $\Lambda$CDM model.

Specifically, we have taken into account some broken power law
$f(R)$ models fulfilling the main cosmological requirements which
are to match the today observed accelerated expansion and the
correct behavior in early epochs. In principle, the adopted
parameterization allows to fit data at extragalactic and
cosmological scales \cite{Hu}. Furthermore, such models are
constructed to evade the Solar System experimental tests. Beside
these broken power laws, we have considered also two models
capable of reproducing the effective cosmological constant, the
early inflation and the late acceleration epochs \cite{sebast}.
These $f(R)$-functions  are combinations of hyperbolic tangents.

We have discussed the behavior of all the considered models. In
particular,  the problem of  stability has been addressed
determining suitable and physically consistent ranges of
parameters. Then we have taken into account the results of the
main  Solar System  current experiments. Such results give upper
limits on the PPN parameters which any self-consistent theory of
gravity should satisfy at local scales. Starting from these, we
have selected the $f(R)$ parameters fulfilling the tests. As a
general remark, all the functional forms chosen for $f(R)$ present
sets of parameters capable of matching the two main PPN
quantities, that is $\gamma_{exp}$ and $\beta_{exp}$. This means
that, in principle, extensions of GR are not {\it a priori}
excluded as reasonable candidates for gravity theories. To
construct such extensions, the reconstruction method developed in
\cite{odintsov} may be applied.

The interesting feature, and the main result of this paper, is
that such sets of parameters are not in conflict with bounds
coming from the cosmological stochastic background of GWs. In
particular, some sets of parameters  reproduce quite well both the
PPN upper limits and the constraints on the scalar component
amplitude of GWs.

Far to be definitive, these preliminary results indicate that
self-consistent models could be achieved comparing experimental
data at very different scales without extrapolating results
obtained only at a given scale.

\section*{ACKNOWLEDGEMENTS}
This research is supported by INFN-CSIC bilateral project, by {\it
Azione Integrata Italia-Spagna 2007} (MIUR Prot. No. 464,
13/3/2006) grant and by MCIN (Spain) projects FIS2006-02842 and
PIE2007-50I023. The work by S.N. is supported by Min. of
Education, Science, Sports and Culture of Japan under grant no.
18549001.

\end{document}